\documentclass[aps,prl,groupaddresses,twocolumn]{revtex4-1}
\usepackage{bm}
\pdfoutput=1
\usepackage{amsmath}
\usepackage{amssymb}
\usepackage{graphicx}

\begin{document}

\title[]{Time-of-flight of solitary waves in dry and wet chains of beads: experimental results and phenomenological models}

\author{Ra{\'ul} Labb{\'e}}
\email{raul.labbe@gmail.com}
\affiliation{Laboratorio de Turbulencia, Departamento de F{\'i}sica, Facultad de Ciencia,
Universidad de Santiago de Chile, USACH. Casilla 307, Correo 2, Santiago, Chile}

\author{Lautaro Vergara}
\email{lautaro.vergara@usach.cl}
\affiliation{Laboratorio de Turbulencia, Departamento de F{\'i}sica, Facultad de Ciencia,
Universidad de Santiago de Chile, USACH. Casilla 307, Correo 2, Santiago, Chile}

\author{Ignacio Olivares}
\email{ignacio.olivares@usach.cl}
\affiliation{Departamento de F{\'i}sica, Facultad de Ciencia, Universidad de Santiago de Chile, USACH. Casilla 307, Correo 2, Santiago, Chile}

\date{\today}

\begin{abstract}
\noindent

\noindent
A solitary wave is generated by impacting a dry chain of beads on one of its ends. Its speed depends on the speed $v_0$ of the striker and the details of the contact force. The time-of-flight (ToF) of the wave was measured as a function of $v_0$, along with the effect of adding a fluid around the contact points. The ToF displays a complex dependence on the fluid's rheological properties not seen in previous works. A power-law dependence of the ToF on $v_0$ in both, dry and wet cases was found. It turned out that the Hertz plus viscoelastic interactions are not enough to account for our results. Two phenomenological models providing a unified and accurate account of our results were developed.

\end{abstract}


\maketitle

The mechanical properties of the contact between solid spheres are of considerable interest in applications involving high stresses on solid surfaces. An extensively used approach to tackle this problem consists in studying the propagation of solitary waves through chains of beads made of the material of interest \cite{Nes83,LazNes85,CosFalFau97,DarNesHer05}. The interaction of one of the ends of the chain with an external medium has been proven useful as a diagnostic tool of anomalies in that medium \cite{khaDarRiz08,YanSilCla11}. Given that the time-of-flight (ToF)---the travel time of a pulse through the whole chain---depends on the individual interactions taking place between the beads during the pulse propagation, we expect a clear dependence of this global magnitude on the properties of those interactions. Measurements of wave propagation and ToF in 1D chains and 2D lattices under different conditions of the contact force were reported by Coste and Gilles \cite{CosGil99,GilCos03}. In addition, measurements of the ToF of solitary waves in dry and wet chains have been performed previously by Herbold \textit{et al.} \cite{HerNesChi06}  and Job \textit{et al.} \cite{JobSanTap08}. In this work, we have made a detailed study of the effect that the amplitude of the incoming pulse has on the ToF in a dry chain, and the changes produced when the interaction force between the beads is slightly modified by a small volume of oil deposited on, and around the beads' points of contact. To this end, an automated device, capable of controlling the amplitude of the wave with nearly arbitrary resolution, was developed. Our results show that the speed of the solitary wave displays a much richer behavior than merely an increase in the propagation speed found in previous works. In addition, these results suggest that the methods used here might be useful to test theoretical models of the contact forces between beads, including the modifications introduced by the addition of foreign materials---fluids in the present case---on the contact points. In our specific case, a study of the propagation of pulses in a chain of stainless steel beads was made, aimed to unveil the effect of the presence of a small volume of oil on the points of contact between the spheres. To assess the effect of a change in the rheological properties of the fluid, three different types of oil were used. To complete our work, we have developed two phenomenological models, based on well stablished physics principles, that accurately describe the behavior of the waves generated in each system.
\begin{figure}[t]
\centering \vspace{0.3cm} \hspace{-0.2 cm}
\includegraphics[width=.48\textwidth]{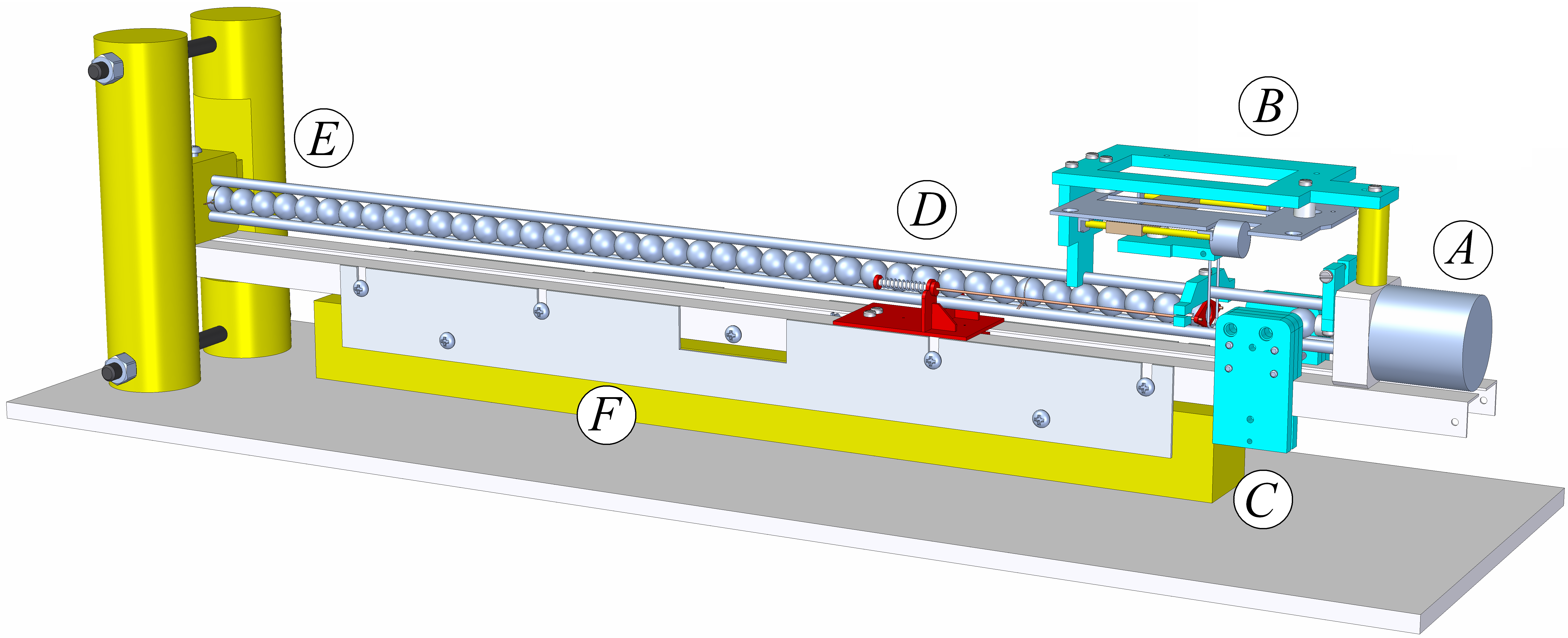}
\caption{Drawing of the experimental setup, showing the mechanical, and electromechanical components. The voice coil motor (A) gives a programmed impulse to the striker ball after it is positioned by the repositioning system (B). Its speed is measured by means of a photogate assembly (C). The spring assembly (D) sets the chain pre-compression force. Electrical signals are generated by piezoceramics at the ball in the site $n=7$ and at the stop at the end wall assembly (E). Two vertical brass rods increase it mass to $4.5$~kg. A second brass bar of mass $6$~kg (F) was added to reduce the mechanical noise transmitted to the chain when the striker is fired. (Details can be seen by zooming into the online figure).}
\label{Fig_1}
\end{figure}

In several aspects our experimental setup is similar to those used in the already mentioned works. A drawing is displayed in Figure \ref{Fig_1}. We use $40 $ stainless steel spheres of type 316, with Young modulus 213 GPa, Poisson ratio $\nu = 0.3$, mass $m = 8.54$ g and radius $R = 6.35$ mm. An electromagnetic actuator is used to give a determined impulse to a ball used as the striker, which impacts the bead at the beginning of the chain and generates a solitary wave. The chain ends with a half of a bead fixed to the flat face of a steel stop, with a Piezosystem's piezoceramic disk in-between. The stop is fixed to a brass structure with mass $4.5$~kg. The arriving mechanical pulses are detected through the charge provided by the ceramic disc, which is converted into a voltage signal by a charge amplifier. A split ball containing a piezoceramic disk between its hemispheres detects the passage of the wave at the site $n = 7$ of the chain. The time elapsed from this event to the arrival of the wave to the end of the chain defines the ToF. To avoid gaps between the beads, a device with two soft springs gently pushes the first bead against the chain with an adjustable force $F \approx 0.24$~N. In addition, the track has a slight inclination of less than one degree, to obtain an idle contact force---as small as possible---between the beads. The striker speeds can be varied from some $0.05$~m/s through about $2.5$~m/s, with a theoretical resolution of about $1.5 \times 10^{-4}$~m/s. Given that the signals from both transducers do not overlap in time, they were added into a single channel to maximize the time resolution of the acquired data.  A photogate allows for precise measurement of the striker speed just before the impact with the chain. All the parameters of this device can be set by a program running in a PC, which also runs the experiments and acquires the data. Automation was necessary to acquire the large amount of data required for statistical purposes. All of the data generated after each shot was acquired through a $16$ bit ADC interface working at $500$~kS/s per channel, or one channel at $1.0$~MS/s, when a better time resolution was needed. The four experiment runs reported here (the dry chain, and wet chains using three types of oil) provided a total of $4.8\times 10^8$ data points. A description in full detail of the experimental device will be published elsewhere.

\begin{figure}[t]
\centering \vspace{0.3cm} \hspace{-0.2 cm}
\includegraphics[width=.48\textwidth]{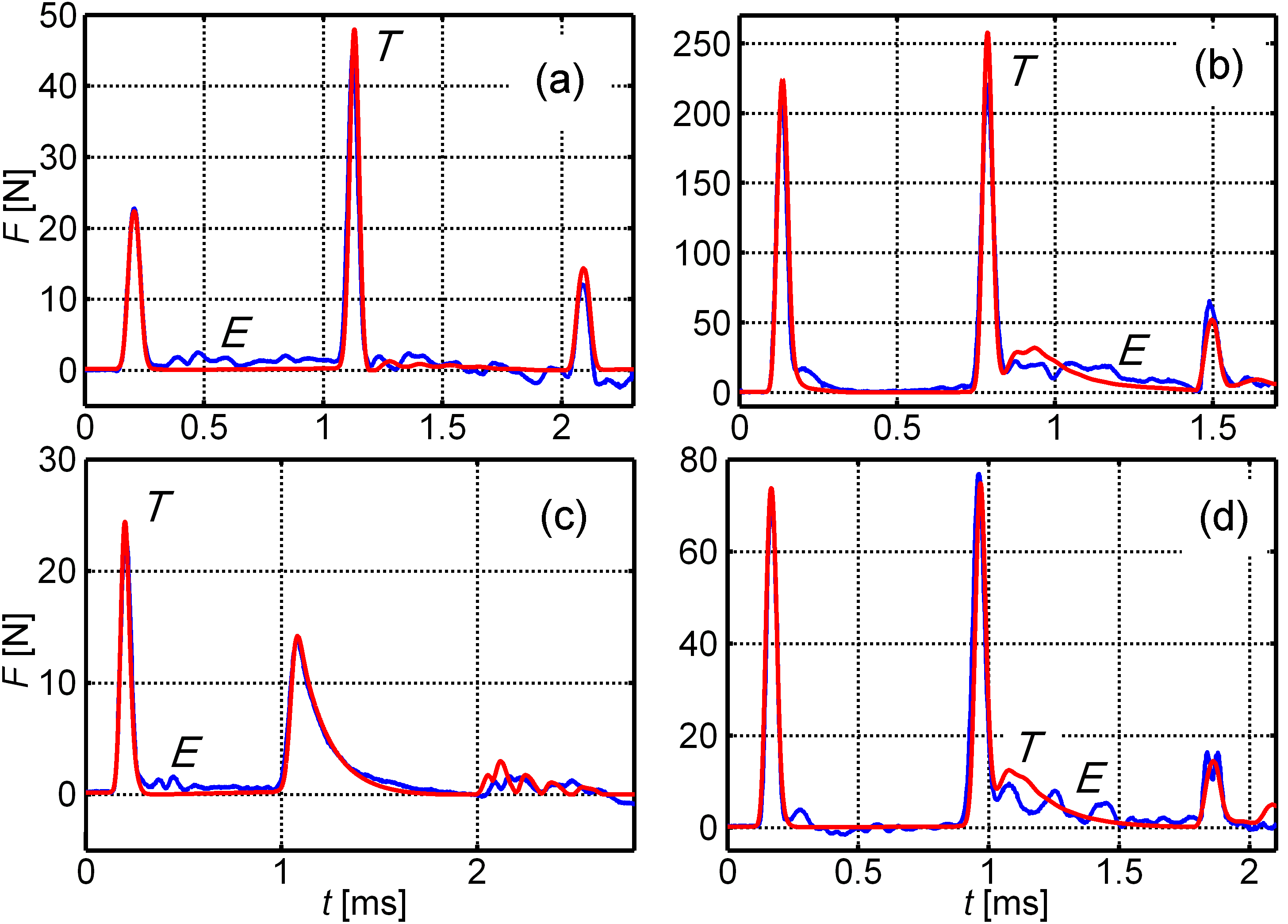}
\caption{Signals (E, blue) from the piezoceramic sensors when the impactor speed is (a) $0.1368$~m/s (dry), (b) $1.1520$~m/s (DS19), (c) $0.1396$~m/s (EP 80W-90), and (d) $0.3993$~m/s (SAE 0W-30). The curves obtained from simulations based on our model (T, red) are drawn on the same plots (see text).}
\label{Fig_2}
\end{figure}

\begin{figure}[t]
\centering \vspace{-0.35cm} \hspace{-0.0 cm}
\includegraphics[width=8.6cm,height=5.8cm]{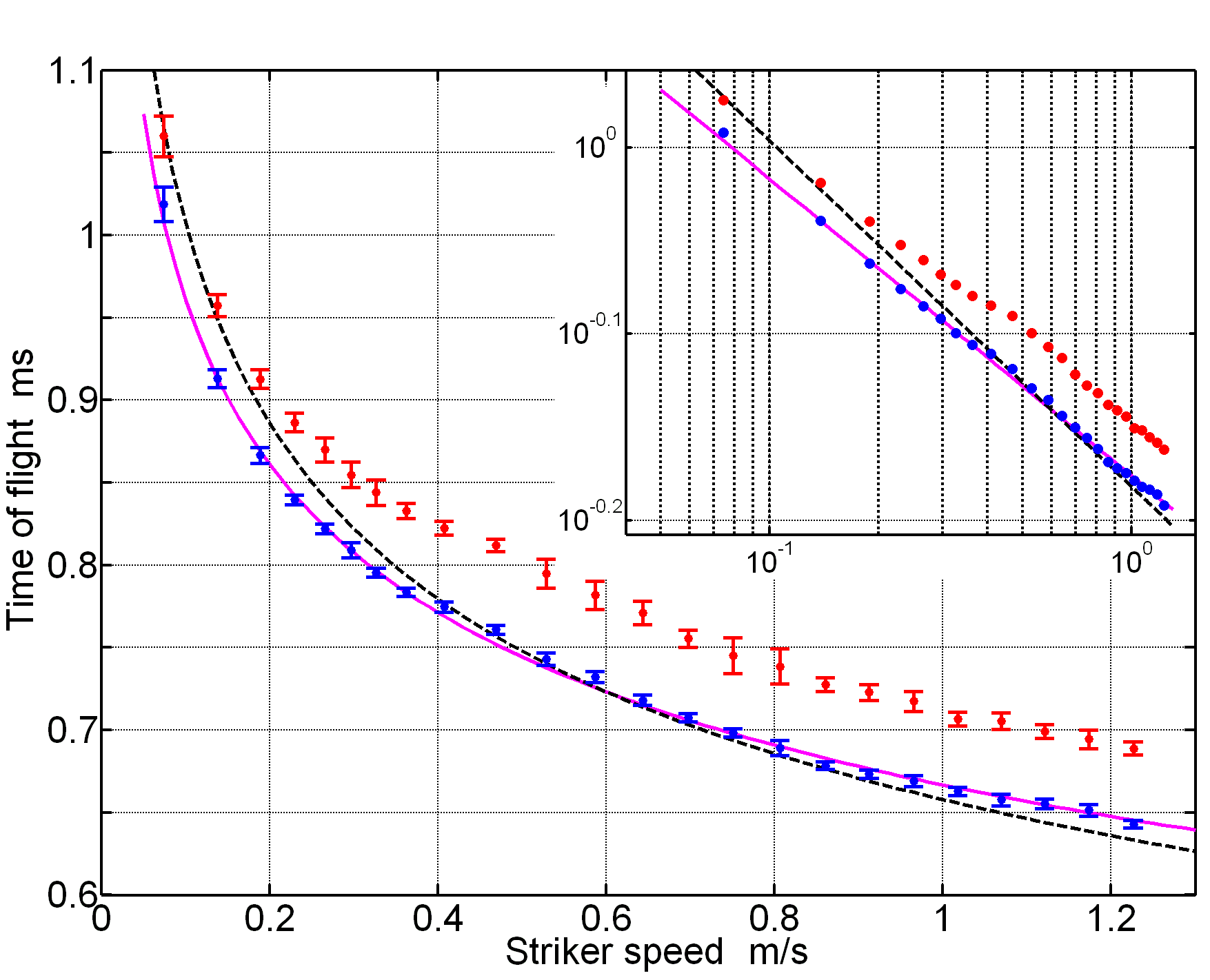}
\caption{ToF in a dry chain made of stainless steel beads, as a function of the striker speed. The lower points correspond to the ToF of incident waves, while the upper points are the ToF of reflected waves off the wall. The continuous curve is the output of our model. The dashed curve is the output of a model with only Hertz plus viscoelastic terms. The inset displays the same data in a log-log plot. Note that both, the ToF and uncertainties are larger for the upper data due to fragmentation of the chain (see text).}
\label{Fig_3}
\end{figure}

\begin{figure}[t]
\centering \vspace{0.0cm} \hspace{-0.2cm}
\includegraphics[width=8.6cm,height=5.7cm]{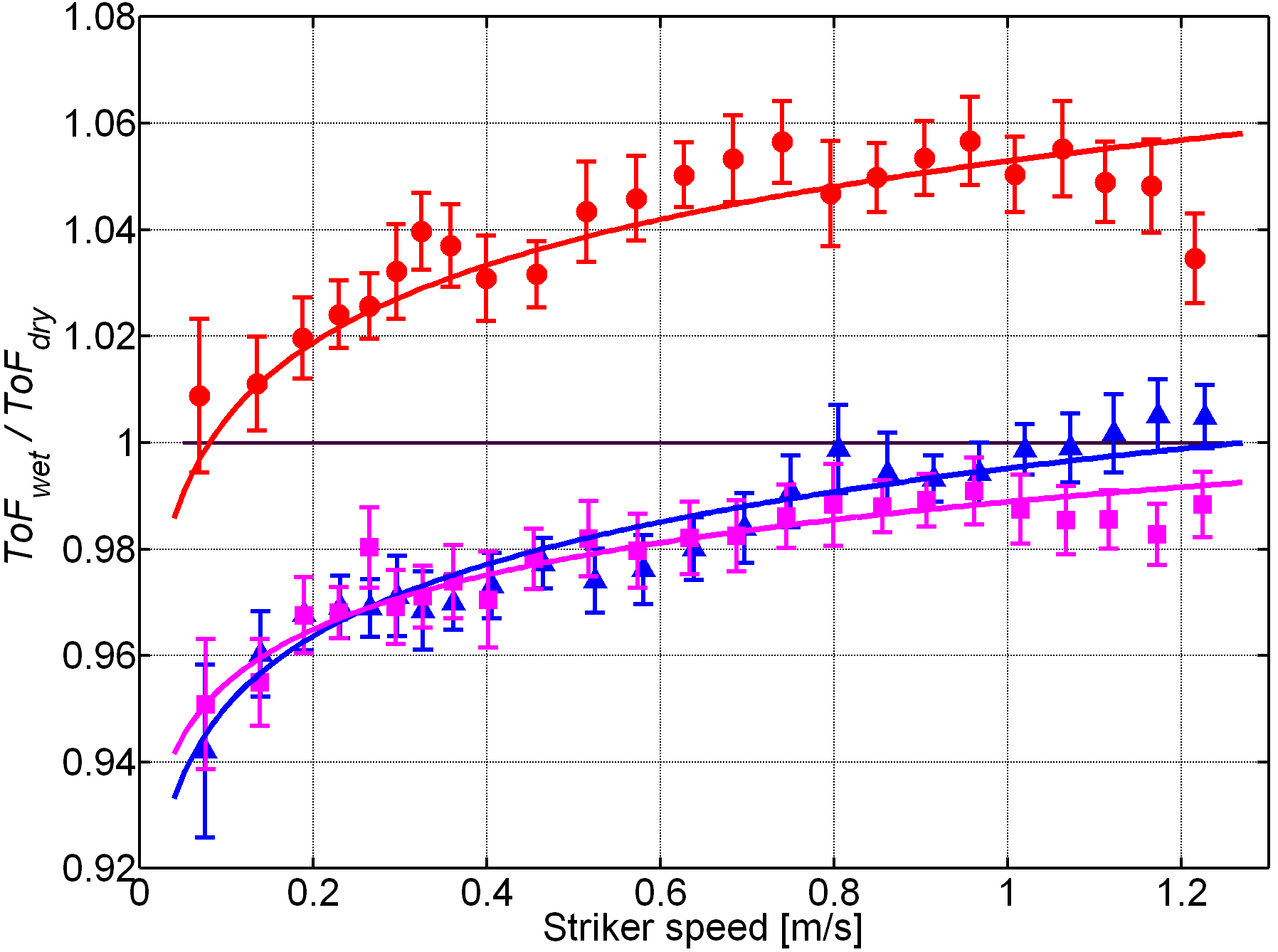}
\caption{Ratios of ToF in wet chains to that in a dry chain. Triangles correspond to rotary vacuum pump oil, squares correspond to gear oil, and filled circles correspond to synthetic motor oil. Continuous curves are the ratios between outcomes from our model for the respective cases.  (see text).}
\label{Fig_4}
\end{figure}

The experimental results shown here were obtained using striker speeds from $0.075 \pm 0.003$~m/s through $1.23 \pm 0.003$~m/s, where the uncertainty was calculated as the standard deviation over $20$ shots. At moderately high speeds, it was observed that at equal impact speeds the rebound speeds of the striker were noticeably different. Subsequently, small motions of the beads along the chain were detected. This suggested that tiny gaps between the beads were generated after the passage of the solitary wave. We performed some numerical simulations, which confirmed that observation. In fact, several authors have studied the fragmentation, and the effects of dispersion, on a line of balls \cite{HerrSch81,HerrSei82,HinSaiJ99}. Thus, a change in the methodology was in order. At speeds below $\sim 0.25$~m/s, no rebound of the striker was observed. Then, before each measurement the striker was used to give a series of taps to the chain, using a speed $v_t=0.18$~m/s. In principle, this should compactify the chain. To make sure that no gaps remained in the chain, prior to each measurement $40$ taps at $v=v_t$ were given. Of course, this made the time necessary for an experiment $40$ times longer than an experiment without tapping. It must be remarked that a tapping technique was previously used by several other authors to increase the packing density of granular materials \cite{KniFanLau95,PhiBid02}. Summarizing, the plots in figures \ref{Fig_2} through \ref{Fig_4} were obtained by repeating $20$ times the sequence of giving $40$ taps, and then shooting the striker at the desired speed, for each point or curve.

In Figure \ref{Fig_2}, four curves represent the signals generated by the piezoceramic transducers for the dry and three wet cases. Theoretical curves overlap almost completely the peaks of the experimental ones. The first pulse on each curve correspond to the incoming wave passing through the ball at the site $n = 7$, the pulses in the middle are produced by the wall sensor, while the last pulses correspond to the reflected wave passing again through the site $n = 7$ in the opposite direction. The curve (a) corresponds to the dry chain. The other curves were obtained with different types of oil between the beads: (b) Varian rotary vane fluid
DS19 type, kinematic viscosity $\nu_k=55 $ cSt, density $\rho=0.870$ g/ml; (c) Castrol, EP 80W-90 gear oil, $\nu_k=133$ cSt, $\rho=0.898$ g/ml, and (d) Castrol, SAE 0W-30 synthetic motor oil, $\nu_k=68.5$ cSt, $\rho=0.8426$ g/ml. $\nu_k$ measured at 40$^{\circ}$C and $\rho$ at 15$^{\circ}$C. Therefore the dynamic viscosities for these three oils are aproximately 0.048 Pa s, 0.119 Pa s and 0.058 Pa s, respectively. It can be seen that the presence of the fluid increases the attenuation of the wave, and modifies the time of flight. The last plot illustrates a case of splitting of the reflected pulse due to the fragmentation of the chain after the transmission of the forward pulse

In Figure \ref{Fig_3} a plot of the ToF on the dry chain is displayed. The lower data correspond to the solitary wave traveling towards the fixed wall, while the upper data represent the ToF of the reflected wave. As expected, the ToF is larger in the latter case. It is also clear that the uncertainty of the data is greater for the reflected wave. These two effects are explained as follows: when the wave is traveling towards the wall, the propagation takes place on a chain compacted by the tapping process. But the wave leaves behind a loosened chain, which is the medium where the reflected wave propagates. This increases both, the ToF and its uncertainty. Without loosening, the increase in the ToF of the reflected pulse would be smaller. The lower data shows clearly the dependence of the ToF on the speed of the striker. Of course, the amplitude of the solitary wave is expected to be a strictly increasing function of the striker's speed, which is confirmed by the experimental data: higher impactor speeds lead to larger wave amplitudes, which in turn lead to faster propagation speeds and smaller ToFs. The continuous curve is the output of our model. The dashed curve is the ToF resulting from a simulation using the Hertz plus viscoelastic interaction force between the beads. The inset shows that this result also follows a power-law dependence of the ToF on the striker speed, although it does not fit the experimental result. In what follows, the lower data are used as the reference to which the data obtained on wet chains is compared.

Figure \ref{Fig_4} displays the effect of adding a drop of oil on the contact points of the beads between the sensor at site $n=7$ and the wall. Triangles correspond to DS19 rotary vacuum pump oil; squares to EP 80W-90 gear oil, and circles to SAE 0W-30 synthetic motor oil. As can be seen, the ratio $T_\mathrm{wet}/T_\mathrm{dry}$ for DS19 oil is less than one for impactor speeds below $1.1$~m/s, meaning that the ToF is smaller than that on the dry chain or, equivalently, that the propagation speed of the solitary wave is larger. For impactor speeds greater than $1.1$~m/s we see the opposite. Although there are some regions where the ToF decreases while the impactor speed increases, the trend of the data is mainly increasing. The data represented in squares was obtained with gear oil, and follow a similar trend, except that at higher impactor speeds the ratio $T_\mathrm{wet}/T_\mathrm{dry}$ remains below one. In this case, the propagation speed is always larger than that in the dry chain. Interestingly, the behavior with motor oil is very different. As the filled circles show, the ToF is greater than one, meaning that the propagation speed in this case is always smaller, but the general trend somewhat resembles the behavior of the gear oil. Notice that although the dynamic viscosities of DS19 oil and SAE 0W-30 motor oil are similar, $T_\mathrm{wet}/T_\mathrm{dry}$ are rather different. On the contrary the dynamic viscosity of EP 80W-90 gear oil is 2.5 times that of DS19, but $T_\mathrm{wet}/T_\mathrm{dry}$ follows almost the same trend. This shows that the dynamic viscosity does not plays a simple role in the dynamics of nonlinear waves in wet chains.

Another remarkable finding is that the four data sets for ToF follow power laws. The continuous curves correspond to the ratios of the wet cases to the dry one, and clearly the curves are in excellent agreement with the experimental data. Thus, as mentioned earlier, our results reveal a complex behavior, not seen in previous works. This shows that the developed models captures the essence of the dynamic of waves on all of the cases. Of course it would be interesting to have full models based on first principles, but this is a far more difficult task. A theoretical model incorporating a detailed knowledge of the rheological properties of the fluids is required, beyond previous efforts to elucidate the effects of viscosity and inertia in the flow between flat surfaces \cite{WeiLawKua85,LawKuaWei85}. Probably the behaviors of the motor and gear oils, as compared to DS19 oil, are related to some non Newtonian characteristic conferred by additives used in these oils to improve their performance. Thus, measurements of ToF in bead chains under appropriate conditions could make them useful as instruments to measure the rheological properties of fluids.\\

Previous attempts to model the dynamics of highly-nonlinear solitary waves in a dry chain of beads were not completely successful \cite{CarKhaPor09,Ver10}. The first one, because it is somehow {\it{ad-hoc}}. The second one was made from a reduced experimental data set, so it is not able to describe the dynamics of the present, larger data set.

For the dry case our model combines the Hertz theory, a viscoelastic term recently obtained by D.S. Goldobin et al. \cite{GolSusPimBri15}, both well based from a theoretical point of view, and a phenomenological force proportional to the third power of the relative velocities of beads.

Let $x_{i}(t)$ represent the displacement of the center of the $i$-th sphere, of mass $m_i$, from its initial equilibrium position. The equations of motion that describe the dynamics of $N$ beads, inclined by an angle $\alpha$, in a gravitational field are:

\begin{widetext}
\begin{equation}
m\ddot{x}_{i}=K_{i-1,i}\, \delta_{i-1}^{3/2}-K_{i,i+1}\,\delta _{i}^{3/2}+\,\frac{3}{2}\Big\{ A_{i-1,i}\,K_{i-1,i}\,\sqrt{\delta _{i-1}}\,\dot{\delta}_{i-1}-A_{i,i+1}\, K_{i,i+1}\,\sqrt{\delta_{i}}\,\dot{\delta}_{i}\Big\}+\,B\left\{\dot{\delta}_{i-1}^{3}-\dot{\delta}_{i}^{3}\right\}+\,m\,g\,\sin \alpha,  \label{uno}
\end{equation}
\end{widetext}
with $i=2,...,N-1$, except for the first and last beads, as known. The overlap between adjacent beads is $\delta _{i}=\max \{\Delta _{i,i+1}-(x_{i+1}-x_{i}),0\}$, ensuring that the spheres interact only when in contact. $\dot{\delta}_{i}=\dot{x}_{i}-\dot{x}_{i+1}$ is the relative velocity of beads, being constant when beads are not in contact. $\Delta _{i,i+1}=(g\,\sin \alpha \,i\,m_{i}/K_{i,i+1})^{2/3}$ appears from the gravitational pre-compression. The expression for the Hertz coupling $K_{i,j}$ between beads $i$ and $j$ is well known to depend on the Young modulus, $Y$, Poisson ratio, $\nu$, and the radius of the beads, $R$. The coupling $A_{i,j}$ is proportional to $(4/3) \eta_{1} (1-\nu+\nu^2)+\eta_{2} (1-2 \nu)^2$, where $\eta_{1}$ and $\eta_{2}$ are the shear and bulk viscosity coefficients, respectively. We follow \cite{ZheZhoYu13} where it is shown that these two coefficients are of the same order of magnitude. Here we simply assume that both are equal. The term cubic in velocities is phenomenological, being $B$ a free parameter with units kg s$/\text{m}^2$.

The set of Eqs. (\ref{uno}) is solved by using a 4th order Runge-Kutta method. With $\eta_{1} = \eta_{2} = 3.2 \times 10^4$ Pa s, $B = 77.0 $ kg s$/\text{m}^2$ this model reproduces quite well the dynamics of nonlinear waves in the present system, including the form of the forward and backward solitary waves, as well as its behavior at the reflecting wall together with the time-of-flight, as shown in Figure \ref{Fig_4}. It is observed that the ratio $\eta/Y = 1.5 \times 10^{-7}$ $\text{s}^{-1}$ is similar to the value $10^{-6}$ $\text{s}^{-1}$ obtained in \cite{ZheZhoYu13} as a result of a finite element calculation for a material with $Y = 10$ GPa and $\nu = 0.3$. As far as we know, no values of viscosity coefficients for stainless steel have been previously measured or simulated.

Modeling the dynamics of highly-nonlinear waves in wet chains from first principles is a difficult, still unsolved problem. Previous attempts \cite{HerNesChi06} and  \cite{JobSanTap08} to introduce some physical input have given at most just a qualitative description of the observed phenomena.

In this paper, for the wet case we considered a simple model with Hertz and viscoelastic terms only. We have assumed that the Young modulus of the system takes an effective constant value on the wet section, different for each oil, and made the Poisson ratio and viscosity coefficient to depend on the impact velocity. Given the complicated dependence of the viscoelastic term on Poisson ratio, in principle such a choice would lead to different, unrelated values for each quantity. However, we have found a surprising power-law behavior, depending on the impact speed, for both Poisson ratio and viscosity coefficient. Our results for that dependence are shown in Table I.

\begin{table}[]
\begin{tabular}{|l|l|l|l|}
\hline
       & Motor                     & Gear                      & Pump                     \\ \hline
Y [GPa]      & 193                       & 225                       & 216                      \\ \hline
$\eta$ [Pa s] & 66177.1 $v_{0}^{-0.206}$ & 107299.0 $v_{0}^{-0.640}$ & 81405.3 $v_{0}^{-0.302}$ \\ \hline
$\nu$  & 0.360 $v_{0}^{-0.165}$   & 0.390 $v_{0}^{-0.127}$   & 0.314 $v_{0}^{-0.274}$  \\ \hline
\end{tabular}
\caption {Power law behavior for Poisson ratio and coefficients of viscosity as a function of the impact speed, $v_0$}
\end{table}

In conclusion, high precision experiments involving $20$ repetitions per value have have been performed in dry and wet chains of beads to study the dynamics of nonlinear waves in these systems, in particular the time-of-flight. A complex behavior, not observed previously, was found: nonlinear waves in wet chains can increase or reduce their speed, with respect to the dry case, depending on the type of oils and on the amplitude of the impact that generate them. With the help of that data we have been able to develop two physics-based models, giving excellent agreement with the experiment.

The authors gratefully acknowledge C. Daraio and P. Anzel for sharing their experience on the use of piezoceramic transducers, and D.S. Goldobin for sending them his work. Financial support for this work was provided by FONDECYT grants No. 1130492, and No. 1040291.


%

\end{document}